\documentclass{wow3_iss_1}
\usepackage{graphicx}
\usepackage{amsmath}
\usepackage{epsfig}
\def\cut#1{}
\def\bm#1{\mbox{\bf \boldmath $ #1 $}} 
\def\rr{\rm I\kern-.2em R} \def\m@th{\mathsurround=0pt} 
\def\ialign{\everycr={}\tabskip=0pt \halign} \def\eqnalign#1{\null 
\vcenter{\openup1\jot \m@th 
\ialign{\strut\hfil$\displaystyle{##}$&$\displaystyle{{}##}$\hfil 
\crcr#1\crcr}}} 

\title[SONAR Visualisation and the Dissimilarity Space]{Analysis of multibeam SONAR data using ICA and the dissimilarity space}
\author{Iain Rice$^\dagger$, $^\ddagger$Roger Benton, $^\ddagger$Les Hart and $^\dagger$David Lowe}
\affiliation{$^\dagger$Aston University, UK. $^\ddagger$Thales Underwater Systems Limited, UK.}

\begin{document}

\maketitle

\begin{abstract}
This paper considers the problem of low-dimensional visualisation of very high dimensional information sources for the purpose of situation awareness in the maritime environment. In response to the requirement for human decision support aids to reduce information overload (and specifically, data amenable to inter-point relative similarity measures) appropriate to the below-water maritime domain, we are investigating a preliminary prototype topographic visualisation model. The focus of the current paper is on the mathematical problem of exploiting a relative dissimilarity representation of signals in a visual informatics mapping model, driven by real-world sonar systems. An independent source model is used to analyse the sonar beams from which a simple probabilistic input model to represent uncertainty is mapped to a latent visualisation space where data uncertainty can be accommodated. The use of euclidean and non-euclidean measures are used and the motivation for future use of non-euclidean measures is made. Concepts are illustrated using a simulated 64 beam weak SNR dataset with realistic sonar targets.

%This paper considers the problem of low-dimensional visualisation of very high dimensional information sources for the purpose of situation awareness in the maritime environment. In response to the
%requirement for human decision support aids to reduce information overload and so to represent high dimensional
%structured data (specifically, data amenable to inter-point relative similarity measures) appropriate to the below-water
%maritime domain, we are investigating a preliminary prototype topographic visualisation model. The focus of the current paper is
%on the mathematical problem of exploiting a relative dissimilarity representation of signals in a visual informatics mapping model, driven by real-world sonar systems. An independent source model is used to analyse the sonar beams from which a simple probabilistic input model to represent uncertainty is mapped to a latent visualisation space where data uncertainty can be accommodated. The use of euclidean and non-euclidean measures are used and the motivation for future use of non-euclidean measures is made. Concepts are illustrated using a simulated 64 beam weak SnR dataset with realistic sonar targets.
\end{abstract}

\section[Introduction]{Introduction} %KEEP IT SHORT OTHERWISE THE STYLE FILE BREAKS????

SONAR (sound navigation and ranging) is used extensively in underwater acoustics. The fundamental issues commonly encountered in the SONAR domain include low signal-to-noise ratio, multipath reflections from targets, sea surface and floor and the high volume of data for analysis. Large sonar systems can produce hundreds to thousands of
beams worth of data, the display of which can easily result in data overload for sonar operators when presented in conventional lofargram displays. Considering the multi-beam time series signals as a sequence of time-dependent vectors in a high dimensional observation space (where the dimensionality is the number of beams being recorded), and this observation sequence is the result of an unknown latent generative model consisting of multiple signals mixed with noise, the desire is to find a low-dimensional latent space representation of the high dimensional data which preserves structure in the original data which would be useful to an operator. Such `topographic' (structure-preserving) projections need to be generically nonlinear and ideally capable of handling the mapping projection of new data without being `re-trained' for each block of new data.  Additionally, since the observation data is always subject to uncertainty, if there is a framework and information to represent uncertainty in the observation data, the projection method should accommodate the uncertainty in the latent space construct.  Finally, since each observation vector (sonar beam scan) only has significance in the context of its relative information content compared to the context of a large number of other scans (ie, human operators perceptually `integrate' lofargrams to smooth out noise, spot anomalous behaviour, occluded tracks etc) we are interested in methods which use relative measures of dissimilarity between pairs of observations, rather than assuming the observed values of a single isolated observation vector are the fundamental entities. In this paper we introduce one such nonlinear topographic visualisation approach which is flexible enough to use different metrics for representing `similarity' between observations vectors, which need not be metric or even positive definite, and can also use measures of `distance' between distribution functions.

\section[Neuroscale]{\label{ns_section}NeuroScale: Topographic dimensionality reduction}

We seek a dimension-reducing, \emph{topographic} transformation of
data for the purposes of visualisation and analysis.  By
`topographic', it is  implied that the geometric structure as determined by pairwise relationships of the data 
is preserved in the transformation. This is a requirement that
relative `dissimilarities' are preserved on average, if possible.

The NeuroScale approach employs a nonlinear transformation $\{\bm{f}:\rr^{n} \rightarrow 
\rr^{m}\ :\  \bm{f}(\bm{x})=\bm{y}\}$ 
from the original configuration space that maps into the feature space. 
 We choose the class of nonlinear parameterised transformations provided 
by Radial Basis Function networks. 
This has the advantage that a {\em transformation\/}  is obtained, 
allowing interpolation. 
  The model parameters are adjusted to minimise the
global \textsc{STRESS} analogous to the classical multidimensional scaling method, (but now with the advantage that a transformation is provided and not just a mapping): 
$$ \eqnalign{
E=\sum_{p=1}^{P}\sum_{q<p}\left[d_{n}(\bm{x}_{p},\bm{x}_{q})-d_{m}(\bm{y}_{p},\bm{y}_{q})  \right]^{2},                 
}
$$
where $d_{n}(\bm{x}_{p},\bm{x}_{q})$ is the `distance' between data 
points in 
the original space (often, but not necessarily, taken to be a euclidean distance between points $||\bm{x}_{p}-\bm{x}_{q}||$) and $\{d_{m}(\bm{y}_{p},\bm{y}_{q})=||\bm{y}_{p}-\bm{y}_{q}||\}$ are 
the distances  in the latent visualisation space (which can also be relaxed if prior knowledge dictates otherwise). The visualisation space dimensionality is often taken to be $m=2$, but $m=1$ and $m=3$ dimensions can also be useful.
The points $\bm{y}$ are generated by the Radial Basis Function network,
 given the data points as
input. That is, $\bm{y}_q = \bm{f}(\bm{x}_q; \bm{\theta})$, where $\bm{f}$ is
the nonlinear transformation effected by the Radial Basis Function model with parameters,
i.e. weights and any kernel smoothing factors, $\bm{\theta}$. The
 (squared) `distances' in the feature space (assuming euclidean discrepancies) may thus be
given by 
%\footnote{it is usual to employ a Euclidean metric space for the projected feature space although this %may also be relaxed if necessary.}
$$
\eqnalign{
  d_{m}^{2} &  = \parallel \bm{f}(\bm{y}_q) - \bm{f}(\bm{y}_p)
 \parallel ^{2}\cr
{}&=\sum_{l=1}^m \left( \sum_k \lambda_{lk} \left[ \phi_k(\parallel
      \! \bm{x}_q \! - \! \bm{\mu}_k \! \parallel) -
      \phi_k(\parallel\!  \bm{x}_p \! -\!  \bm{\mu}_k \!
      \parallel) \right] \right) ^2 .\cr
}
$$

The topographic nature of the transformation is imposed by the \textsl{STRESS}
term which attempts to match the inter-point dissimilarities in
the latent visualisation space with the dissimilarities in the input space. Note that nowhere is an isolated point vector needed; only measures or pairwise dissimilarity.

Also, note that central to this transformation is the assumption of a provided `distance' functions $d_{n},d_{m}$ in input and latent spaces, and also the choice of interpolating basis spline functions, $\phi(\dots )$. 

\vspace{1pc}\subsection[Dissimilarity Measure]{Dissimilarity measure}

A priori, there is no reason to suppose that the dynamical evolution in the observation or latent spaces should occur on a Euclidean manifold (zero curvature). Prior knowledge on sensors for example may indicate a geometry of input data space different from euclidean.  So using a euclidean measure of distance to characterise dissimilarity between vectors in the input space may be an incorrect simplification. The Bregman divergence is a class of measures which allows for incorporating a degree of `manifold curvature' into measures of similarity. The Bregman divergence between two points $\bm{p}, \bm{q}$ can be expressed as
\begin{equation}
d_{F}(\textbf{p},\textbf{q}) = F(\textbf{p}) - F(\textbf{q}) - \langle (\textbf{p}-\textbf{q}), \nabla F(\textbf{q})\rangle
\end{equation}
where $F(.)$ represents a pre-defined function and $\langle \dots \rangle$ represents the inner product.
The positivity of Bregman divergence as a measure of `distance' stems from the constraint of the convexity of $F()$. The Bregman divergence measures the difference between the manifold value at the point $\bm{p}$ with a first order Taylor expansion about the point $\bm{q}$ and hence reflects the departure due to the curvature of the manifold. However, note that generally the Bregman divergence is not a metric in that it does not satisfy the triangle inequality and so is not a true `distance'.  For different choices of convex function, $F()$, different measures may be derived, such as the squared Euclidean dissimilarity ($F(\bm{X}) = \langle X,X\rangle$), KL divergence (if $\bm{X}$ is a discrete probability distribution such that $\sum_{i=1}^{N} X_{i} =1$ then choose $F(\bm{X}) = \sum_{i=1}^{N}X_{i} \log_{2} X_{i}$), Bayes Risk error (choose $F(\bm{X}) = -J(\bm{X})$ where $J(\bm{X})$ is the Bayes risk of the optimal detection rule), and so on.  

In an interesting twist, a recent extension to the standard stress function in topographic mappings was given in \cite{wang11} using Bregman divergences with the choice of $F(x) = x \log (x)$ as the convex function measuring the deviations between the dissimilarities in the input space and the discrepancies in the latent visualisation space.

\subsection[Non-positive definite splines]{Non-positive definite splines}

It is usual for smoothing splines used in interpolation to be positive definite.  The natural domain of analysis of positive kernel smoothing splines is reproducing kernel Hilbert spaces where the vector space is endowed with a positive definite inner product.  However in several data analysis situations it is not intrinsic to the problem that the inner product should necessarily be positive. The radial basis function used in the Neuroscale model can deploy non positive definite metrics and basis functions. The natural domain of splines with non-positive definite metrics is a reproducing kernel Krein space~[\cite{canu05,hassibi96}].  This leads to pseudo-Riemannian manifold descriptions of data rather than the archetypal euclidean manifold.  In \cite{ting12} a Minkowski metric was used in a problem where a lorentzian manifold was appropriate for the problem domain.  Due to space, this generalisation to the sonar data is the result of future work.

\section[Results]{Results}

\subsection[Data]{Data}
\begin{figure}
%\vspace{5pc}
\includegraphics[width = 7cm]{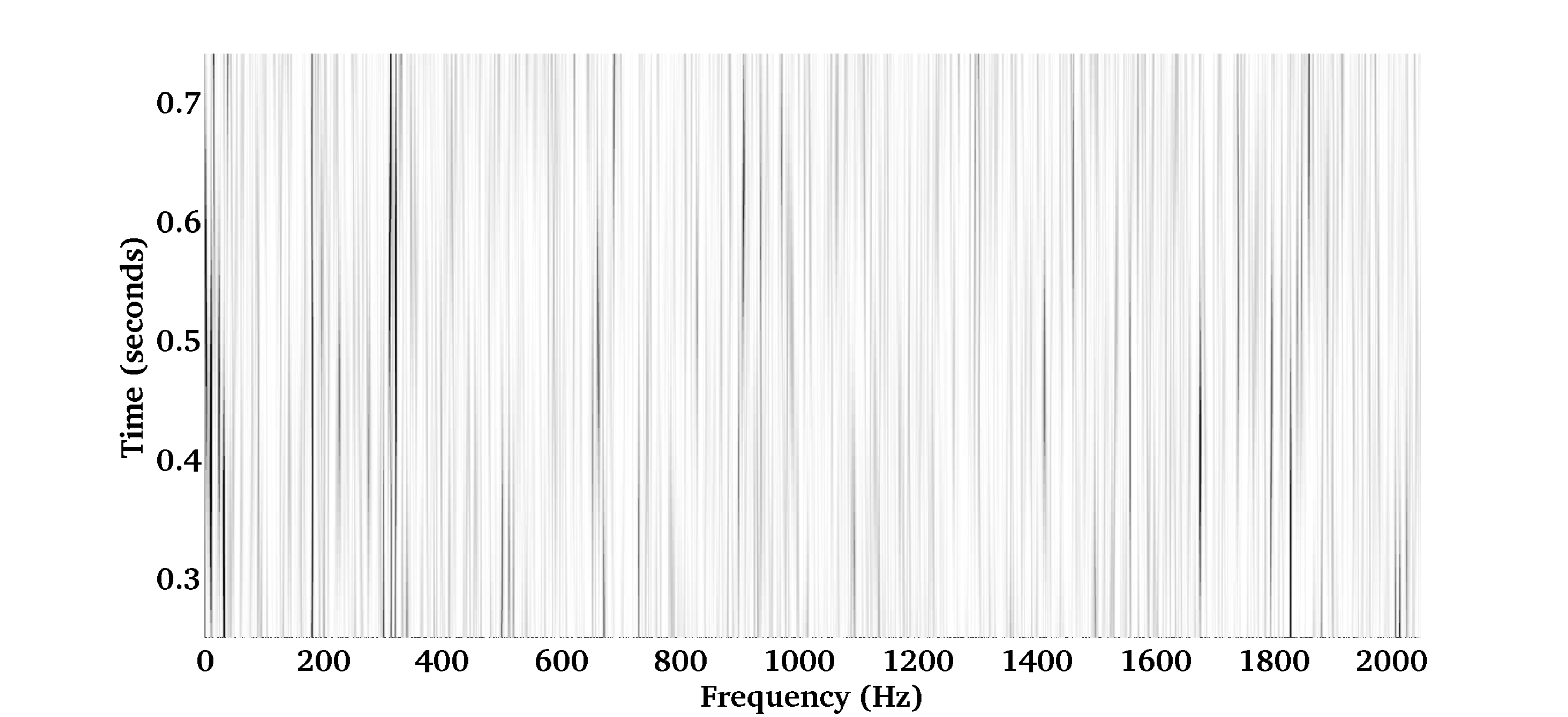}
\includegraphics[width = 7cm]{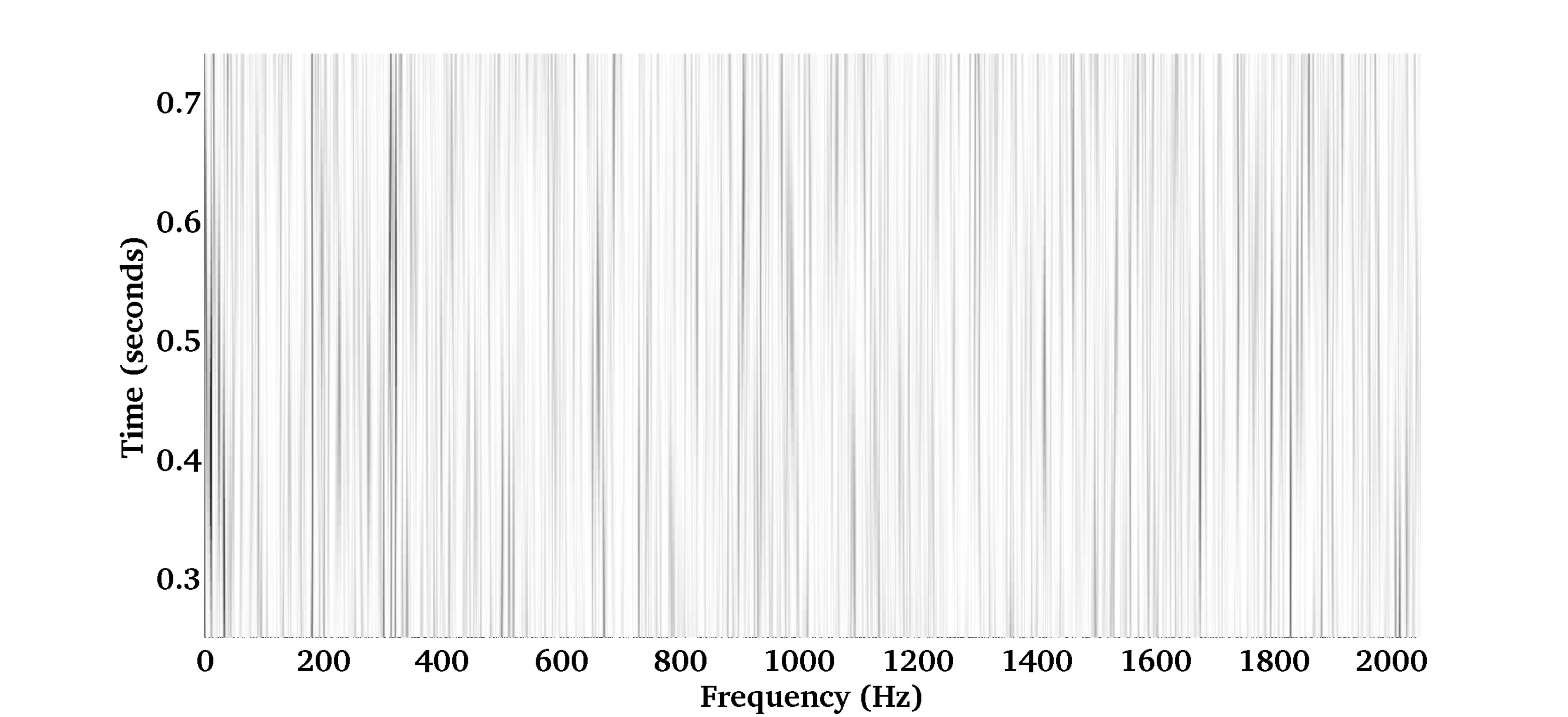}
\caption{\label{lofargram}Lofargram showing 1 second of the multibeam simulated data, Left subfigure is the original data, RIght subfigure shows the resonstructed `cleaned' data observed through the ICA embedding model where prominent target data remains and some background noise activity is removed. }
\end{figure}

A simulated typically weak SnR 64-beam SONAR dataset has been used for preliminary analysis. The dataset has realistic targets with additive white Gaussian noise. The sampling rate is 4096 Hz. A typical lofargram of a subset of the data is shown in figure~\ref{lofargram}. 
There are 3 prominent targets that move through several beams over the full 488 second simulation. 

\subsection[Time Series Analysis]{Time Series Analysis}

Although not key to this paper, prior to dissimilarity pattern processing, the raw data was subject to a `filtering' to remove the obvious noise subspace.  The method we used followed the approach given in ~\cite{lowe98,lowe01,woon04} where we assume a dynamical systems perspective: that the scalar observations are the result of an observation function on a latent higher dimensional state vector evolving on an unknown manifold, but subject to observational and dynamical noise. Using the embedding approach as described in ~\cite{broomhead86} to recreate a topologically-equivalent manifold space and then the approach in~\cite{lowe98,woon04,lowe01} we extracted a set of common latent sonar `sources' using the single channel ICA approach. The set of sources was automatically split into a set describing the noise subspace, and the remaining set describing the signal subspace. Then the observed sonar signals were reconstructed using only the signal sources. This approach is equivalent to an FIR filtering of the data where the filter coefficients are data-driven.  The sources were obtained on a large representative data set and were subsequently held fixed for all future signal processing.  The results in this paper were produced using these ICA-subspace filtered signals. 

\subsection[Exemplar 3D Projection]{Exemplar 3D Projection}

%IAIN.. CHECK THE WORKDING HERE.. I HAVE HAD TO GUESS A LOT OF WHAT YOU ACTUALLY DID.. ALSO WITH REGARD TO THE OUTLIER DATA POINTS IN THE VIS SPACE.. THEY SHOULD BE BACK-REFERENCED TO DATA IN THE LOFARGRAM REALLY, BUT I GUIESS WE DONT HAVE TOIME TO DO THIS..SO I HAVE SURMISED WHAT THEY ARE LIKELY TO BE !!!!

Each processed temporal slice of 64 time series samples is treated as a dynamical point observation vector in a 64D space which needs to be transformed into a 2D or 3D visualisation space. Each point is regarded as the mean of a gaussian distribution in the observation space, where the (time-dependent and spherical) covariance of the Gaussian is estimated using the variance across all beams as a proxy for uncertainty information.  This is merely for illustration of the method being applied to a distribution.  Typically, different measures of uncertainty would be available through other means. 

\begin{figure}
%\vspace{5pc}
\includegraphics[width = 7.5cm]{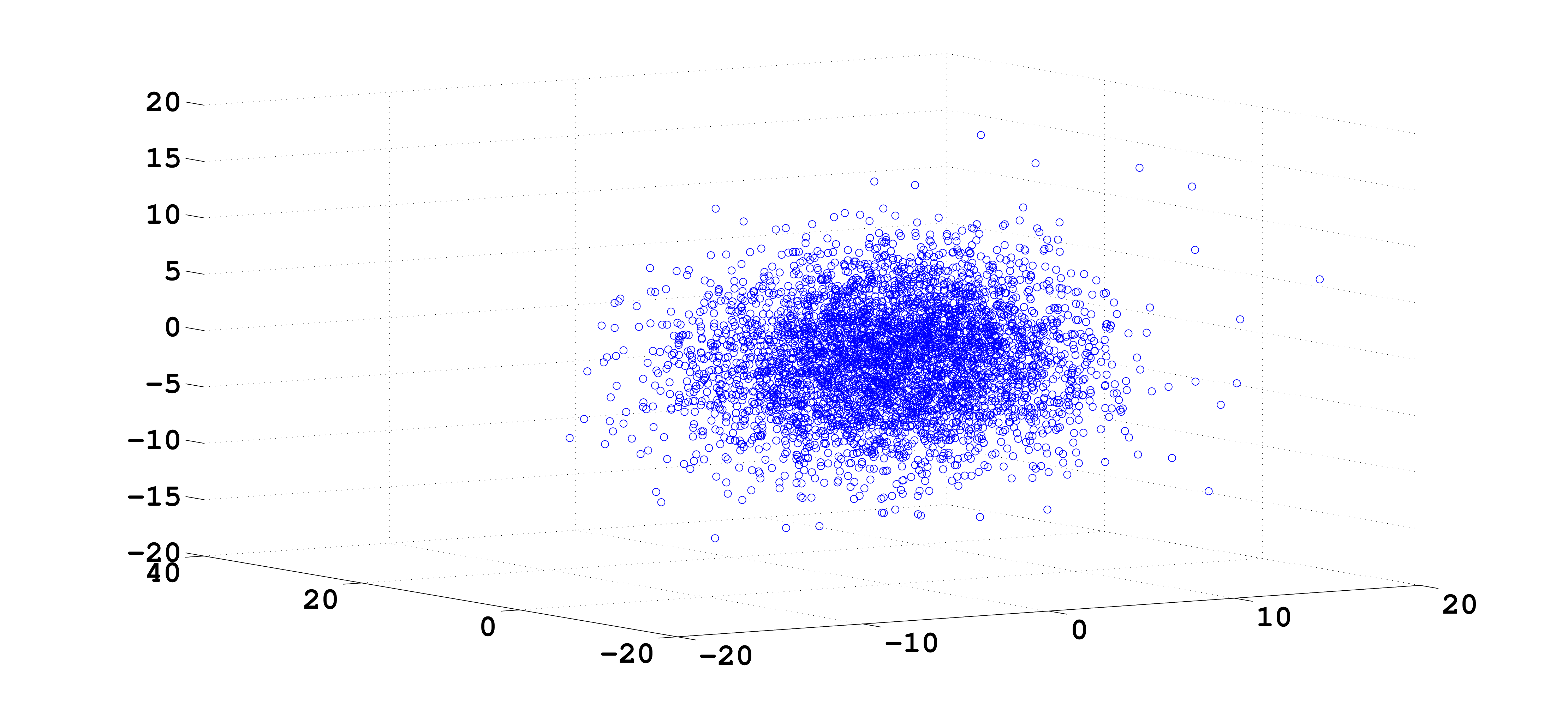}
\includegraphics[width = 7.5cm]{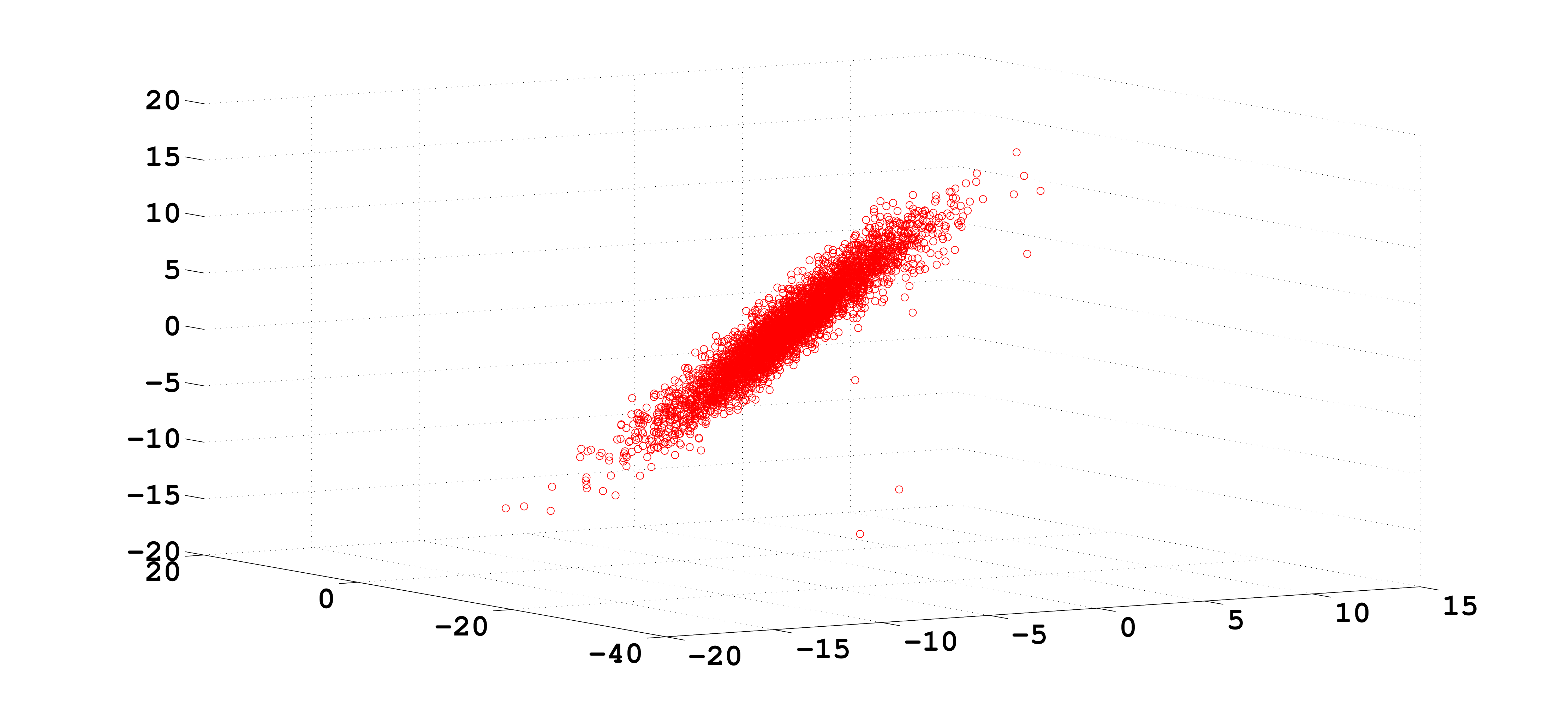}
\caption{\label{vis1}Euclidean (left) and non-Euclidean (right) comparisons of the Neuroscale 3D visualisation of 4096, 64D data samples. The Euclidean visualisation is more prone to outliers, and the `interesting' data points are more apparent in the non-Euclidean visualisation.}
\end{figure}

Following the discusssion in \ref{ns_section}, the left-hand subfigure in Figure~\ref{vis1} shows a 3D neuroscale scatterplot visualisation of 1 second of sonar data at full temporal resolution (4096 observations, each of 64 dimensions) using euclidean dissimilarity. The right-hand subfigure in Figure~\ref{vis1} shows the same data segment using a non-euclidean visualisation. The euclidean neuroscale visualisation uses the approach in \cite{ting12} to map points and their associated uncertainties. The non-euclidean visualisation uses one-sided KL divergences between distributions in input and visualisation spaces and Bregman divergence to replace the STRESS measure.  The non-euclidean visualisation of data and its uncertainty is less prone to outliers and also highlights specific 'interesting' data points where target data of particular note can be observed. (These points reflect target signatures in the original data which represent major departures from the `normality' of surrounding data and hence would be highlighted to the operator for analysis.)

\subsection[Beam Clustering]{Beam Clustering}

\begin{figure}
%\vspace{5pc}
\centering{\includegraphics[width = 10cm]{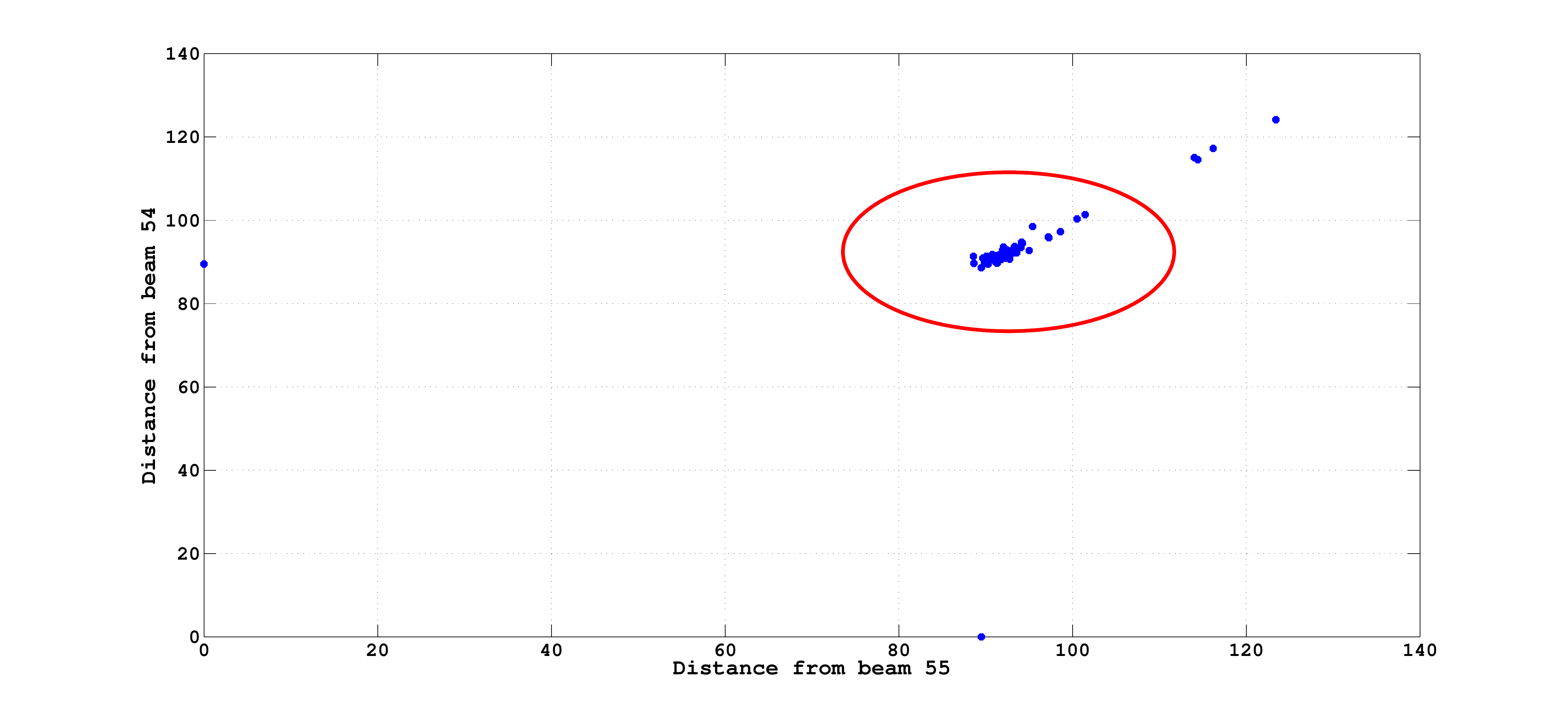}}
\caption{\label{cluster1}
3D PSD dissimilarity  of the 64 channels measured relative to three prototypes, channels 62, 55 and 56. Most channels are clustered in one region with small dissimilarity. Points of significantly different dissimilarity represent the actual targets which are located in these channels. The points located on the axes correspond to channels 62, 55 and 56.
}
\end{figure}

To highlight the `interesting' beams, we now consider a more traditional use of the dissimilarity representation, following the approach in \cite{duin00,duin09} for dissimilarity clustering. For this same segment of data we now consider the power spectrum of each channel of data (after the temporal ICA-processing to remove obvious noise subspaces), and  measure the dissimilarity between each channel's spectrum relative to a set of prototype channels.  These prototypes were determined automatically using a modeseek algorithm to select three reference channels (channels 62, 55 and 56 happened to be selected) against which the dissimilarities were measured. This set of 64, 3D points is given in Figure~\ref{cluster1}.
Figure~\ref{cluster1} indicates that a cluster of beams is
present where many beams are situated within a small range of dissimilarity between the
two representatives. A smaller cluster of points with larger dissimilarity is present
containing beams 1,2,32 and 33. The 3 targets present in the simulated data are centred
at these beams 1 and 2, 32 and 33 and in the two representatives 54 and 55.

\vspace{2pc}\section[Conclusion]{Conclusions}

Taking a dissimilarity view of pattern processing has significant  potential for improving the ability of operators to analyse high dimensional and poor quality sonar data, both through providing augmented low dimensional visualisation representations to the conventional lofargram display, and through automated alarm systems to automatically highlight data regions of significant departure from `normal' behaviours. However this requires an investigation into appropriate domain-specific measures of dissimilarity. In this paper we have introduced a dissimilarity-based visuaisation approach which allows the exploration of a range of interesting and nonconventional measures, some of which need not be distances at all, and others that can compare between distributions.  This latter ability is important in maritime situations in which data is categorised not just by point samples, but augmented by some measure of uncertainty or importance of a given datapoint. 

%\cite[Hwang \& Tuck 1970]{Hwang70}; \cite[Lee 1971]{Lee71}).

%\begin{figure}
%  \vspace{5pc}
%  %\includegraphics[width=8cm]{fig1.eps}
%  \caption{Shaded contour plots of the potential $\phi$ for the two trapped %modes that exist for an ellipse with $a/d=1.5$, $b/d=0.75$.    (\textit{a}) %Symmetric about $x=0$, $kd=0.96$; (\textit{b}) antisymmetric about $x=0$, %$kd=1.398$.}
%  \label{fig:contour}
%\end{figure}


\begin{thebibliography}{}

\cut{	\bibitem[Belouchrani et. al. (1993)]{belouchrani93} \textsc{Belouchrani, A., Abed-meraim, K., Cardoso, J. F. and Moulines, E.} 1993 Second Order Blind Separation of Correlated Sources. \textit{Proc. Int. Conf. on Digital Signal Processing}, 346--351.
	\bibitem[Belouchrani \& Cichocki (2000)]{sobi01} \textsc{Belouchrani, A. and Cichocki, A.} 2000 SOBI MATLAB algorithm [Online]. Available: http://sccn.ucsd.edu/eeglab/allfunctions/sobi.m.
}
		
		\bibitem[Broomhead \& King (1986)]{broomhead86}\textsc{Broomhead, D. S. and King, G. P.} 1986 Extracting qualitative dynamics from experimental data. \textit{J. Phys. D} \textbf{20}, 217--236.
	

\bibitem[Canu (2005)]{canu05} \textsc{Canu S,  Ong CS, and Xavier M},
2005
{Splines with Non Positive Kernels},
in \textit{5th International ISAAC Congress} World Scientific.

\bibitem[Hassibi et al (1996)]{hassibi96} \textsc{Hassibi B, Sayed A.H. and Kailath T} 1996 
{Linear estimation in Krein spaces - part I: Theory}
 \textsc{IEEE Transactions on Automatic Control}, \textbf{41}(1) 18--33. 	
	
\cut{	\bibitem[Cichocki \& Amari (2002)]{cichocki02} \textsc{Cichocki, A. and Amari, S.} 2002 Adaptive Blind Signal and Image Processing: Learning Algorithms and Applications.
}
\cut{	\bibitem[Davies \& James (2007)]{davies07} \textsc{Davies, M. E. and James, C. J.} 2007 Source separation using single channel ICA. \textit{Signal Processing} \textbf{87, 8}, 1819--1832.
}
	\bibitem[Duin (2000)]{duin00} \textsc{Duin, R.P.W.} 2000 Classifiers in almost empty spaces. \textit{Proc. 15th Int. Conf. on Pattern Recognition, 2000.} \textbf{2}, 1--7.
		\bibitem[Duin \& Pekalska (2009)]{duin09} \textsc{Duin, R.P.W. and Pekalska, E.} 2009 Datasets and tools for dissimilarity analysis in pattern recognition. \textit{SIMBAD Project, Information and Communication Technologies Collaborative Project} \textbf{3.3}.
			
		
	\cut{	\bibitem[FastICA (2005)]{fastica05} \textsc{G\"{a}vert, H., Hurri, J., S\"{a}rel\"{a}, J. and  Hyv\"{a}rinen, A.} 2005 FastICA MATLAB algorithm [Online]. Available: http://research.ics.aalto.fi/ica/fastica/.
}
		
\cut{					\bibitem[Ghosh \& Unnikrishnan (2005)]{ghosh05} \textsc{Ghosh, S. and Unnikrishnan, A.} 2005 Implementation of Independent Component Analysis on SONAR signals using Neural Networks. \textit{2nd International Conference on Electrical Engineering/Electronics, Computer, Telecommunications and Information Technology (ECTI-CON 2005)} \textbf{1}, 558--561.
	}		
					
	\cut{			\bibitem[Givens (1958)]{givens01} \textsc{Givens, W.} 1958 Computation of Plane Unitary Rotations Transforming a General Matrix to Triangular Form. \textit{J. Soc. for Industrial and Applied Mathematics} \textbf{6, 1}, 26--50.
		}
		\cut{		\bibitem[Hyvarinen \& Oja (1997)]{hyvarinen97} \textsc{Hyv\"{a}rinen, A. and Oja, E.} 1997 A Fast Fixed-Point Algorithm for Independent Component Analysis. \textit{Neural Computation} \textbf{9}, 1483--1492.
		}
		\cut{	\bibitem[Hyvarinen (1999)]{hyvarinen99} \textsc{Hyv\"{a}rinen, A.} 1999 The Fixed-Point Algorithm and Maximum Likelihood Estimation for Independent Component Analysis. \textit{Neural Processing Letters}, 1--5.
	}
	
	\cut{		\bibitem[Kamalet.al.  (2011)]{kamal11} \textsc{Kamal, S. and Supriya, M. H. and Pillai, P.R.S.} 2011 Mitigating ambient noise in underwater acoustic receivers using independent component analysis. \textit{2011 Int. Symposium on Ocean Electronics (SYMPOL)}, 184--193.
}

\cut{	\bibitem[Koldovsk\'{y} \& Tichavsky (2011)]{koldovsky11} \textsc{Koldovsk\'{y}, Z. and Tichavsky, P.} 2011 Time-Domain Blind Separation of Audio Sources on the Basis of a Complete ICA Decomposition of an Observation Space. \textit{Audio, Speech, and Language Processing, IEEE Transactions on} \textbf{19, 2}, 406--416.
}

\bibitem[Lowe (1998)]{lowe98} \textsc{Lowe, D.} 1998 Feature space embeddings for extracting structure from single channel wake EEG using RBF networks, {\textit Neural Networks for Signal Processing VIII, 1998. Proceedings of the 1998 IEEE Signal Processing Society Workshop}, 428--437.

			\bibitem[James \& Lowe (2001)]{lowe01} \textsc{James, C. and Lowe, D.} 2001 Single channel analysis of electromagnetic brain signals through ICA in a dynamical systems framework. \textit{Engineering in Medicine and Biology Society, 2001. Proceedings of the 23rd Annual International Conference of the IEEE} \textbf{2}, 1974--1977.
	
\cut{	\bibitem[de Moura et. al. (2007)]{moura07} \textsc{de Moura, N.N.,  Seixas, J.M.,  Filho, W.S. and Greco, A.V.} 2007 Independent Component Analysis for Optimal Passive Sonar Signal Detection. \textit{7th Int. Conf. on Intelligent Systems Design and Applications, 2007.}, 671--678.
}

\cut{		\bibitem[Naik \& Kumar (2001)]{naik01} \textsc{Naik, G. and Kumar, D.} 2001 An overview of independent component analysis and its applications. \textit{Informatica} \textbf{35}, 63--81.
}

					\bibitem[Pekalska \& Duin(2005)]{pekalska05} \textsc{P\c{e}kalska, E. and Duin, R.P.W.} 2005 The Dissimilarity Representation For Pattern Recognition. \textit{World Scientific}.

				\bibitem[Pekalska et. al. (2006)]{pekalska06} \textsc{P\c{e}kalska, E.,Duin, R.P.W. and Paclik, P.} 2006 Prototype Selection for Dissimilarity-based Classifiers. \textit{Pattern Recognition} \textbf{39, 2}, 189--208.

\cut{
\bibitem[Pourkhaatoun et. al. (2007)]{pour07}  \textsc{Pourkhaatoun, M., (Reza) Zekavat, Seyed A. and Pourrostam, J.} 2007 A Novel High Resolution ICA Based TOA Estimation Technique. \textit{Radar Conference, 2007 IEEE}, 320-324.
}

\cut{	\bibitem[Takens (1981)]{takens81} \textsc{Takens, F.} 1981 Detecting strange attractors in turbulence. \textit{Dynamical Systems and Turbulence, Lecture Notes in Mathematics} \textbf{898}, 366--381.
}

\bibitem[Wang et.al. (2011)]{wang11} \textsc{Wang, X. Fyfe C, and Chen W.} 2011 Applyling Bregman Divergences to the Neuroscale
Algorithm, \textit{Proceedings of the 11th UK Workshop on Computational Intelligence}, 190-195.

\bibitem[Wang \& Lowe (2012)]{ting12}\textsc{Wang, X. and Lowe, D.}  2012 Signal Processing Issues of High--Dimensional Visual Informatics: A
Study in Protein--DNA  Binding Patterns. \textit{9th IMA International Conference on Mathematics in Signal Processing.} 


\bibitem[Woon \& Lowe (2004)]{woon04} \textsc{Woon, W.L., Lowe, D.} 2004 Can we learn anything from single-channel unaveraged MEG data?' {\textit Neural Computing and Applications}, \textbf{13}(4), 360--368.

\end{thebibliography}
\end{document}